\documentclass{article}

\usepackage{times}
\usepackage{enumitem}
\usepackage{latexsym}
\usepackage{complexity}
\usepackage[group-separator={,}]{siunitx}
\usepackage{color}
\usepackage{tikz}
\usetikzlibrary{positioning,chains,fit,shapes,calc}
\usepackage{natbib}

\newtheorem{mytheorem}{Theorem}

\newcommand{\myqed}{\mbox{$\Box$}}
\newcommand{\mylike}{\mbox{\sc Li\-ke}}
\newcommand{\myblike}{\textsc{Ba\-lan\-ced Li\-ke}}

\title{Sustainable Fair Division}
\author{Martin Aleksandrov \\
UNSW Australia and Data61 \\
 martin.aleksandrov@data61.csiro.au}

\begin{document}

\maketitle

\section{Introduction}

Nowadays, many resources on planet Earth are scarce and threatened of extinction. Some of them are even only partially accessible to us. At the same time, the extant resources often arrive online and need to be allocated in highly constrained environments. For this reason, we need a simplistic manner in which we allocate resources online that enables us to satisfy various economic, ecological and societal needs in our daily life. Sustainable resource allocations are vital for the survival of our planet. For example, we work with a food bank allocating donated food to charitable organizations. Recent reports showed that giving out food to people in need improves their family life and their productivity at work\footnote{https://www.foodbank.org.au/}. As another example, we collaborate with a national organ exchange scheme on allocating organs of deceased donors to patients on a waiting list. Now, the allocation decisions could save lives. Evidently, online fair divisions are important due to their significant returns. In this abstract, I summarize our work on online fair division. In particular, I present two models for online fair division: (1) one existing model for fair division in food banks and (2) one new model for fair division of deceased organs to patients. I further discuss simple mechanisms for these models that allocate the resources as they arrive to agents. In practice, agents are often risk-averse having imperfect information. Within this assumption, I report several interesting axiomatic and complexity results for these mechanisms and conclude with future work.   

\section{Foodbank Allocations}

We formulated an online model for fair division in foodbanks. In this model, $m$ indivisible items arrive in some strict ordering $o$ and are allocated to $n$ agents. Each agent has some non-negative (private) utility for each item. Every time an item arrives, the agents report bids for the item thus revealing their valuations for it and a mechanism allocates it to one of the agents. \cite{aleksandrov2015ijcai} provides initial axiomatic analysis of two such mechanisms for this model: (1) \mylike\ gives uniformly at random each next item to an agent that bids positively and (2) \myblike\ gives uniformly at random each next item to an agent with fewest items among those agents that bid positively. As these mechanisms are randomized, each agent receives each item with some \emph{ante probability}. Their \emph{expected utility} is the dot product between the vectors of their ante probabilities and their cardinal utilities for items. I studied axioms of these mechanisms such as {\em envy-freeness ex post}, {\em ex post efficiency} and \emph{competitiveness} among many others. For example, \mylike\ can be much less fair than \myblike. Even with 0/1 utilities, {\sc Like} could give all items to only one agent. In contrast, {\sc Balanced Like} tries to give each agent the same number of items. However, both of these mechanisms are ex post efficient with 0/1 utilities. They satisfy this property because, in each actual allocation, they give out all items only to agents that value them positively and the sum of the agent's ex post utilities is equal to $m$.

\begin{mytheorem}
With 0/1 utilities, {\sc Like} and {\sc Balanced Like} are ex post efficient.
\end{mytheorem} 

Even with utilities in $\lbrace 0,1,2\rbrace$, checking efficiency is computationally expensive. We reduce from the $\coNP$-hard problem {\sc Res\-tri\-cted\-3UN\-SAT}: given a set  $C=\lbrace c_1,\ldots,c_m\rbrace$ of clauses each of at most 3 literals over $n$ propositions such that each of them occurs twice positively and once negatively in $C$, is $C$ unsatisfiable; see \cite{demange2008}?

\begin{mytheorem}
With utilities in $\lbrace 0,1,2\rbrace$ and either mechanism, checking Pareto efficiency of an allocation is $\coNP$-hard.
\end{mytheorem} 

\begin{myproofsketch}
I depict the reduction in terms of example: $P=\lbrace p_1,p_2,p_3\rbrace$, $c_1=\lbrace p_1,p_2,\neg p_3\rbrace$, $c_2=\lbrace \neg p_1,p_2,p_3\rbrace$, $c_3=\lbrace p_1,\neg p_2, p_3\rbrace$. Similarly as in \cite{keijzer2009}, we define allocation $\pi(I)$ for each partial interpration $I$ over the propositions. We next represent the allocation instance.

\begin{center}
\resizebox{\columnwidth}{2cm}{
$\begin{array}{ccccccccccccccccccc}
o  & i_{p_1} & i_{p_2} & i_{p_3} & i_{c_1} & i_{c_2} & i_{c_3} & i_{sat,p_1} & i_{sat,p_2} & i_{sat,p_3} & i_{p_1c_1} & i_{p_2c_1} & i_{\neg p_3c_1} & i_{\neg p_1c_2} & i_{p_2c_2} & i_{p_3c_2} & i_{p_1c_3} & i_{\neg p_2c_3} & i_{p_3c_3} \\
a_{p_1} & 2 & 0 & 0 & 0 & 0 & 0 & 0 & 0 & 0 & \color{black}{\bf [1]} & 0 & 0 & 0 & 0 & 0 & \color{black}{\bf [1]} & 0 & 0 \\
a_{\neg p_1} & 1 & 0 & 0 & 0 & 0 & 0 & 0 & 0 & 0 & 0 & 0 & 0 & \color{black}{\bf [1]} & 0 & 0 & 0 & 0 & 0 \\
a_{p_2} & 0 & 2 & 0 & 0 & 0 & 0 & 0 & 0 & 0 & 0 & \color{black}{\bf [1]} & 0 & 0 & \color{black}{\bf [1]} & 0 & 0 & 0 & 0 \\
a_{\neg p_2} & 0 & 1 & 0 & 0 & 0 & 0 & 0 & 0 & 0 & 0 & 0 & 0 & 0 & 0 & 0 & 0 & \color{black}{\bf [1]} & 0 \\
a_{p_3} & 0 & 0 & 2 & 0 & 0 & 0 & 0 & 0 & 0 & 0 & 0 & 0 & 0 & 0 & \color{black}{\bf [1]} & 0 & 0 & \color{black}{\bf [1]} \\
a_{\neg p_3} & 0 & 0 & 1 & 0 & 0 & 0 & 0 & 0 & 0 & 0 & 0 & \color{black}{\bf [1]} & 0 & 0 & 0 & 0 & 0 & 0 \\
a_{c_1} & 0 & 0 & 0 & \color{black}{\bf [1]} & 0 & 0 & 0 & 0 & 0 & 1 & 1 & 1 & 0 & 0 & 0 & 0 & 0 & 0 \\
a_{c_2} & 0 & 0 & 0 & 0 & \color{black}{\bf [1]} & 0 & 0 & 0 & 0 & 0 & 0 & 0 & 1 & 1 & 1 & 0 & 0 & 0 \\
a_{c_3} & 0 & 0 & 0 & 0 & 0 & \color{black}{\bf [1]} & 0 & 0 & 0 & 0 & 0 & 0 & 0 & 0 & 0 & 1 & 1 & 1 \\
a_{un,p_1} & \color{black}{\bf [1]} & 1 & 1 & 0 & 0 & 0 & 2 & 2 & 2 & 0 & 0 & 0 & 0 & 0 & 0 & 0 & 0 & 0 \\
a_{un,p_2} & 1 & \color{black}{\bf [1]} & 1 & 0 & 0 & 0 & 2 & 2 & 2 & 0 & 0 & 0 & 0 & 0 & 0 & 0 & 0 & 0 \\
a_{un,p_3} & 1 & 1 & \color{black}{\bf [1]} & 0 & 0 & 0 & 2 & 2 & 2 & 0 & 0 & 0 & 0 & 0 & 0 & 0 & 0 & 0 \\
a_{sat,c_1} & 0 & 0 & 0 & 1 & 1 & 1 & \color{black}{\bf [1]} & 1 & 1 & 0 & 0 & 0 & 0 & 0 & 0 & 0 & 0  & 0  \\
a_{sat,c_2} & 0 & 0 & 0 & 1 & 1 & 1 & 1 & \color{black}{\bf [1]} & 1 & 0 & 0 & 0 & 0 & 0 & 0 & 0 & 0  & 0  \\
a_{sat,c_3} & 0 & 0 & 0 & 1 & 1 & 1 & 1 & 1 & \color{black}{\bf [1]} & 0 & 0 & 0 & 0 & 0 & 0 & 0 & 0  & 0  \\
\end{array}$
}
\end{center}

Let $I_{\emptyset}$ be the \emph{null} interpretation; i.e. maps no propositions. We showed that the set $C$ is unsatisfiable iff $\pi(I_{\emptyset})$, denoted as the utilities in brackets, is Pareto efficient. The proof resembles the one from \cite{keijzer2009}. However, our construction gives tighter bound on the complexity of the problem and thus gives a better inside on its possible parameterizations. Moreover, we needed to guarantee that the allocation $\pi(I_{\emptyset})$ occurs with positive probability with each of these mechanisms, i.e. each agent is ``eligible'' for each next item in the ordering.\myqed
\end{myproofsketch}

We studied competitiveness of these mechanisms as well. The \emph{egalitarian welfare} is equal to the minimum agent expected utility. In \cite{aleksandrov2015ijcai}, we computed the competitive ratios of these mechanisms from egalitarian perspective against the optimal (offline) mechanism that knows all the information about the future items. In summary, with 0/1 utilities, the offline competitive ratios of \mylike\ and \myblike\ are at least $n$. With general utilities, \mylike\ is $n$-competitive whereas \myblike\ is not at all.

Another interesting issue with these mechanisms is their complexity of computing ex post and ex ante outcomes. With any of them, we can compute actual discrete allocations in $\mathcal{O}(n\cdot m)$ space and time. Interestingly, \mylike\ has the same complexity of computing ante allocations because each agent gets an item they like with ante probability of one divided by the number of agents that like the item. By comparison, \myblike\ is computationally hard. The reduction is parsimonious from the problem of computing the number of perfect matchings in a given undirected bipartite graph; see. e.g. \cite{dagum1992,saban2013}. 

\begin{mytheorem}
With 0/1 utilities, computing the agent probabilities is in $\P$ with {\sc Like} whereas is in $\#\P$-hard with {\sc Bal. Like}.
\end{mytheorem} 

\begin{myproofsketch}
With {\sc Balanced Like}, I illustrate the reduction on a simple example in which a 3-regular bipartite graph $G$ $=$ $(U,V,E)$ is given: $U$ $=$ $\lbrace u_1,u_2,$ $u_3,u_4\rbrace$,  $V$ $=$ $\lbrace v_1,v_2,$ $v_3,v_4\rbrace$ and $E$ $=$ $\lbrace (u_1,v_1),$ $(u_1,v_2),$ $(u_1,v_4),$ $(u_2,v_2),$ $(u_2,v_3),$ $(u_2,v_4),$ $(u_3,v_1),$ $(u_3,v_3),$ $(u_3,v_4),$ $(u_4,v_1),$ $(u_4,v_2),$ $(u_4,v_3)\rbrace$. The allocation instance is represented in the following tabular form.

\begin{table}[h]
\begin{tabular}{c}
\resizebox{\columnwidth}{2cm}{
$\begin{array}{ccccccccccccccc}
o  & v_1 & v_2 & v_3 & v_4 & u^1_1 & u^2_1 & u^1_2 & u^2_2 & u^1_3 & u^2_3 & u^1_4 & u^2_4 & i_1 & \color{black}{\bf [i_2]} \\
e^1_1 & \color{black}{\bf [1]} & 0 & 0 & 0 & \color{black}{\bf [1]} & \color{black}{\bf [1]} & 0 & 0 & 0 & 0 & 0 & 0 & 0 & 1 \\
e^2_1 & 0 & \color{black}{\bf [1]} & 0 & 0 & \color{black}{\bf [1]} & \color{black}{\bf [1]} & 0 & 0 & 0 & 0 & 0 & 0 & 0 & 1 \\
e^3_1 & 0 & 0 & 0 & \color{black}{\bf [1]} & \color{black}{\bf [1]} & \color{black}{\bf [1]} & 0 & 0 & 0 & 0 & 0 & 0 & 0 & 1 \\
e^1_2 & 0 & 1 & 0 & 0 & 0 & 0 & 1 & 1 & 0 & 0 & 0 & 0 & 0 & 1 \\
e^2_2 & 0 & 0 & 1 & 0 & 0 & 0 & 1 & 1 & 0 & 0 & 0 & 0 & 0 & 1 \\
e^3_2 & 0 & 0 & 0 & 1 & 0 & 0 & 1 & 1 & 0 & 0 & 0 & 0 & 0 & 1 \\
e^1_3 & 1 & 0 & 0 & 0 & 0 & 0 & 0 & 0 & 1 & 1 & 0 & 0 & 0 & 1 \\
e^2_3 & 0 & 0 & 1 & 0 & 0 & 0 & 0 & 0 & 1 & 1 & 0 & 0 & 0 & 1 \\
e^3_3 & 0 & 0 & 0 & 1 & 0 & 0 & 0 & 0 & 1 & 1 & 0 & 0 & 0 & 1 \\
e^1_4 & 1 & 0 & 0 & 0 & 0 & 0 & 0 & 0 & 0 & 0 & 1 & 1 & 0 & 1 \\
e^2_4 & 0 & 1 & 0 & 0 & 0 & 0 & 0 & 0 & 0 & 0 & 1 & 1 & 0 & 1 \\
e^3_4 & 0 & 0 & 1 & 0 & 0 & 0 & 0 & 0 & 0 & 0 & 1 & 1 & 0 & 1 \\
\color{black}{\bf [s]} & 0 & 0 & 0 & 0 & 0 & 0 & 0 & 0 & 0 & 0 & 0 & 0 & 1 & \color{black}{\bf [1]} \\
\end{array}$
}
\end{tabular}
\end{table}

Intuitively, each agent of each triplet of consecutive agents possibly receives exactly one of the first $3\cdot n$ items; e.g. the utilities of the first three such agents are denoted in bold brackets. With this argument, we showed that there is 1-to-$2^n$ correspondence between the perfect matchings in $G$ and the possible allocations of the first $3\cdot n+1$ items such that each agent gets exactly one item. For this reason, the probability of agent $s$ for item $i_2$ is $\frac{1}{3\cdot n+1}\cdot\frac{1}{3^n}\cdot |\mbox{\sc Perf}(G)|$.\myqed
\end{myproofsketch}

Finally, in \cite{aleksandrov2015ijcai}, I also validated these mechanisms on both generated and real-world preference profiles: uniform binary utilities, correlated binary utilities, Borda utilities, etc.

\section{Deceased Organ Matchings}

There are $n$ patients on a strict waiting list. Each patient has two major compatibility indicators: blood type (O, A, B, AB) and estimated post-transplant survival index (EPTS). At a given moment, a pair of kidneys arrives. Each kidney has similar parameters: blood type (O, A, B, AB) and kidney donor profile index (KDPI). Each patient has utility $-|$EPTS$-$KDPI$|$ for each kidney. Patients in the waiting list may arrive or depart. For example, in Australia, a patient is added to this list nearly every \num{15} min. At the same time, patients may leave the list in case they say have (offline) exchanged organs from living donors. There are some hard constraints: e.g. patients of type O can be matched to organs of types O only. Further, there are some soft constraints as well: e.g. patients of type AB can be matched to organs of types O, A, B and AB, but are preferred to be matched to organs from type AB. Whilst this matching is possibly efficient w.r.t. blood type, it might not be necessarily efficient w.r.t. the ``life-longevity'' indices. However, in Australia, the waiting list contains patients from any blood type. Moreover, only up to \num{5}\% of the people in the waiting list have blood type AB and around \num{50}\% of the people in this list have blood type O. This indicates a real fairness issue: patients of type AB can access \num{100}\% of the donated deceased organs and patients of type O can access only \num{50}\% of them. 

For this reason, I discuss two simple deterministic mechanisms for allocating kidneys to patients that take into account both the blood type and index compatibility. {\sc HardTypeBestIndexMatch} computes the sub-list of patients in the current waiting list who minimize $|$EPTS$-$KPDI$|$ among those with exact blood type match for the new organ. {\sc SoftTypeBestIndexMatch} computes the sub-list of patients in the current waiting list who minimize $|$EPTS$-$KPDI$|$ among those with compatible blood type match for the new organ. Each of these mechanisms splits the list of eligible patients into four sub-lists, one per blood type, and allocates each kidney to an eligible patient from one of the sub-lists that has the greatest length. Ties between sub-lists of the same length are broken by giving priority to the patients with more compatible blood type for the new organ. Ties between patients are broken using the waiting list in favor of agents who has waited the longest; see e.g. \cite{ur1992}. 

I propose several novel properties of the allocation mechanisms. A mechanism is \emph{blood type efficient} if there is no other allocation in which each agent is matched to an organ of some more compatible blood type. A mechanism is \emph{index efficient} if there is no other allocation in which each agent is matched to an organ with index of higher quality. Interestingly, {\sc HardTypeBestIndexMatch} is blood type efficient because each organ is matched to a patient with exactly the same type. On the other hand, {\sc SoftTypeBestIndexMatch} is not blood type efficient. To see this, just consider two patients of type AB and one patient of type O, all with the same survival index, and an organ of type O. This mechanism gives the organ to the patient of type AB that has waited longest. In regard to index efficiency, both of these mechanisms violate this axiom in general simply because, at the current round, they might match an organ with KDPI of 100 to a patient with EPTS of 1 but, at some later round, another organ could arrive with KDPI of 1 for this patient. However, as the length of the waiting list in practice is quite large, we are motivated to consider a subclass of ``exact'' matching instances in which, for each organ, there is a patient in the list with exact index match.

\begin{mytheorem}
{\sc HardTypeBestIndexMatch} is blood type efficient and {\sc SoftType\-BestIndexMatch} is not. With exact instances, {\sc SoftTypeBestIndexMatch} is index efficient and {\sc HardTypeBestIndexMatch} is not.
\end{mytheorem}  

Additionally, I consider two envy-freeness ex post notions. A patient of type $a$ would \emph{envy} a patient of another type $b$ for a compatible organ of type $c$ if $b\not=c$ and the latter patient is allocated the organ. This envy accounts for the blood type. A mechanism is \emph{blood type envy-free} if no agent envies another w.r.t. their blood types. Envy-freeness could be defined for indices as well. A patient with EPTS index $p$ would \emph{envy} with $|q-s|/(|p-s|+\epsilon)$ another patient with EPTS $q$ for a compatible organ with KDPI $s$ if the latter patient is allocated the organ. The value of $\epsilon$ is tiny. A mechanism is \emph{index envy-free} if no agent envies another w.r.t. their indices.

\begin{mytheorem}\label{thm:envy}
{\sc HardTypeBestIndexMatch} is blood type envy-free and {\sc SoftType\-BestIndexMatch} is not. With exact instances, {\sc SoftTypeBestIndexMatch} is index envy-free and {\sc HardTypeBestIndexMatch} is not.
\end{mytheorem} 

Theorem~\ref{thm:envy} is easy to prove. We can further study bounded envy-freeness as well; see e.g. \cite{aleksandrov2016ijcaidc}. 

\begin{mytheorem}\label{thm:envy2}
With any instances, {\sc SoftTypeBestIndexMatch} is bounded index envy-free with 1 and {\sc HardTypeBestIndexMatch} is not.
\end{mytheorem} 

\begin{myproof}
With {\sc SoftTypeBestIndexMatch}, for each organ, one of the ``eligible'' agents receives it. They do not envy any other agent. Each other ``eligible'' agent envies them with at most 1. Each other non-``eligible'' agent envies only this ``eligible'' agent with at most 1 and none of the other agents.

With {\sc HardTypeBestIndexMatch}, consider two patients of types O and AB and EPTS indices of 100 and 1, respectively, and one organ of type O and KDPI of 1. This mechanism matches the organ to the patient of type O and index of 100. But, then the patient of type AB has envy equal to $\infty$ as $\epsilon$ goes to $0$.\myqed
\end{myproof}

Let us next take a look at the competitiveness of these mechanisms. Recall, we assume that there is a patient in the current list of type $b$ for each new organ of type $b$. The optimal (offline) allocation w.r.t. blood types is the one in which each patient receives an organ of their most compatible blood type. The objective is the number of agents matched to organs of the same blood type. {\sc HardTypeBestIndexMatch} achieves this outcome whereas {\sc SoftTypeBestIndexMatch} is sub-optimal. We could also study competitiveness w.r.t. ``life-longevity'' indices. Now, optimally, each patient is matched to an organ of the closest index. The objective is the number of agents matched to organs of the same quality index. In a special case, {\sc SoftTypeBestIndexMatch} maximizes this outcome whereas {\sc HardTypeBestIndexMatch} is sub-optimal.

\begin{mytheorem}
{\sc HardTypeBestIndexMatch} is blood type optimal and {\sc SoftType\-BestIndexMatch} is $2$-competitive w.r.t. blood types. With exact instances, {\sc SoftTypeBestIndexMatch} is index optimal and {\sc HardTypeBestIndexMatch} is $100\cdot n$-competitive w.r.t. indices.
\end{mytheorem} 

\begin{myproof}
For {\sc SoftTypeBestIndexMatch}, it might mismatch at most half of the organs w.r.t. blood types. This happens when it gives each organ of type $b\not=$ AB to a patient of type AB. There are at most half such cross-matches compared to the offline mechanism.

For {\sc HardTypeBestIndexMatch}, consider the example in the proof of Theorem~\ref{thm:envy2}. This mechanism matches the organ to the patient of type O and index of 100. The optimal decision w.r.t. indices is to match the organ to the patient of type AB and index of 1. The results follow.\myqed
\end{myproof}

Finally, as these deterministic mechanisms output a single ex post allocation, we can compute it in $\mathcal{O}(n\cdot m)$ space and time where $m$ is the number of organs. 

\section{Conclusion}

Agents are often entitled to different shares of the resource. For example, in the Foodbank setting, charities have possibly different feeding abilities. As another example, in the Organ Matching setting, each patient could require one or two kidney transplants. How do we then allocate the resources? Finally, how these mechanisms compare empirically against the established matching techniques used in practice.

\bibliographystyle{plainnat}
\bibliography{submission}

\end{document}